\documentstyle[aps,prl,epsf,floats]{revtex}

\begin{document}
\draft

\twocolumn[\hsize\textwidth\columnwidth\hsize\csname@twocolumnfalse%
\endcsname

\title{The anomalous 0.5 and 0.7 conductance plateaus 
       in quantum point contacts}
\author{Henrik Bruus, Vadim V.\ Cheianov, and Karsten Flensberg}
\address{{\O}rsted Laboratory, Niels Bohr Institute,
Universitetsparken 5, DK-2100 Copenhagen \O, Denmark}
\date{Physica E {\bf 10}, 97, (2001)}
\maketitle
\begin{abstract}
The anomalous 0.5 and 0.7 conductance plateaus in quantum point
contacts in zero magnetic field are analyzed within a phenomenological
model. The model utilizes the Landauer-B\"{u}ttiker formalism and
involves enhanced spin correlations and thermal depopulation of spin
subbands. In particular we can account for the plateau values 0.5 and
0.7, as well as the unusual temperature and magnetic field dependences
of the 0.7 plateau. Finally, the model predicts the possibility of 
coexisting 0.5 and 0.7 plateaus.

\vspace*{3mm}

\end{abstract}

\vspace*{3mm}

]

\section{Introduction}
It has been known and well understood since
1988\cite{vanWees88,Wharam88,vanHouten92} that the dc-conductance $G$
of narrow quantum point contacts and quantum wires (both referred
to as QPCs below) is quantized in units of $G_2= 2\: e^2/h$. During
the past five years an increasing part of the experimental and
theoretical work on QPCs has been devoted to studies of a particular
deviation from this integer quantization known as 0.7 conductance anomaly
\cite{Thomas96,Thomas98,Liang99,Kristensen98a,Kristensen98b,Kristensen00,Thomas00,Reilly00,Berggren98,Spivak,BCF}.
This anomaly is a narrow plateau, or in some cases just a
plateau-like feature appearing in scans of $G$ versus gate
voltage $V_g$ at a value of $G$ which is reduced by a factor 0.7 relative
to the ideal value $G_2$. The 0.7 anomaly has been
recorded in numerous QPC transport experiments (even before it
was noted in 1996, see e.g.\ Ref.~\cite{vanWees88}). Recently
the appearance of an anomalous plateau at the value $0.5 G_2$ was
also reported \cite{Thomas00,Reilly00}.

In this paper we show that many of the experimental findings regarding
the 0.5 and 0.7 anomalous plateaus can be consistently interpreted by
invoking the model of enhanced spin correlations in the QPC, which we
previously put forward to explain the 0.7 plateau \cite{BCF}. 
We emphasize that our model does not rely on the existence of static
polarization, which would be inconsistent with some general theorems
\cite{Lieb}, but rather on effects of dynamical local polarization in
the QPC. 

Already in the first paper\cite{Thomas96} it was pointed
out that due to its magnetic field dependence the 0.7 anomaly may be
related to spontaneous spin polarization of electrons in the QPC.
Theoretical attempts to link the 0.7 anomaly to spontaneous
spin polarization have been made \cite{Berggren98,Spivak}. However, none
of these approaches have explained all of the experimental facts, and
most strikingly they failed to predict the observed  plateau at $0.7\:
G_2$. Also the 0.7 anomaly cannot be explained by impurity
backscattering mechanisms, Luttinger liquid effects or an interplay of
both, the predicted temperature dependence being opposite to the
observed one. 

\section{Summary of experimental facts}
In summarizing experimental data we will mainly refer
to the work of the Cambridge group
\cite{Thomas96,Thomas98,Liang99} and the Copenhagen group
\cite{Kristensen98a,Kristensen98b,Kristensen00} presenting detailed
studies of the magnetic field and temperature dependence of the
0.7 anomaly.

The main experimental features of the 0.7 anomaly are:

(e1) The anomalous plateau is observed in a large variety of QPCs
at a value $G = \gamma\: G_2$, where the
suppression factor $\gamma$ is close to 0.7
\cite{Thomas96,Thomas98,Liang99,Kristensen98a,Kristensen98b,Kristensen00}.

(e2) The temperature dependence is qualitatively the same for all
samples: the anomalous plateau is fully developed in some (device
dependent) temperature range typically above 2~K. With increasing
temperature both the anomalous and the integer plateaus vanish by
thermal smearing, while with decreasing temperature the width of the
anomalous plateau shrinks the conductance approaching $2 G_2$.
\cite{Thomas96,Thomas98,Liang99,Kristensen98a,Kristensen98b,Kristensen00}.

(e3) A detailed study of the temperature dependence of $\gamma$
shows that in the low temperature regime the conductance suppression
has an activated behavior: $1 - \gamma(T) \propto \exp(-T_a/T)$
\cite{Kristensen98b,Kristensen00}.

(e4) The activation temperature $T_a$ is a function of
$V_g$ vanishing at some critical gate voltage $V_g^0$. Close to
$V_g^0$ the dependence of $T_a$ on $V_g$ is well approximated by a
power law \cite{Kristensen98b,Kristensen00},
$T_a \propto (V_g - V_g^0)^{\alpha}$, with $\alpha \approx 2$.

(e5) At a fixed temperature corresponding to a well developed 0.7
plateau, $\gamma$ shows a strong dependence on an
in-plane magnetic field\cite{Thomas96,Thomas98,Liang99}.
With increasing magnetic field $\gamma$ smoothly decreases from 0.7
at $B=0$~T to 0.5 at $B=13$~T.

(e6) Under the same temperature conditions as in (e5)
the 0.7 anomaly depends on the source-drain
bias. The suppression factor $\gamma$ increases smoothly from $\sim 0.7$
at zero bias to $\sim 0.9$ at large bias ($\sim 2$~mV)
\cite{Kristensen00}.

(e7) In some samples the anomalous plateau was reported to
appear at $G=0.5 G_2$ \cite{Thomas00,Reilly00} rather than at
$G=0.7 G_2.$

\section{The phenomenological model}

The ground state of the two-dimensinal electron gas (2DEG) has
been studied extensively using ever more refined methods involving
local density functional theory\cite{Stern73,Stern84} and Monte Carlo
calculations\cite{Ceperley78}. In particular it is argued in the
latter work that at at zero temperature there exist at least three
phases of the 2DEG: an unpolarized fluid ($r_s<13$), a fully spin
polarized fluid ($13<r_s<33$), and a Wigner crystal polarization
($r_s>33$). More recently, also using density functional theory,
the quasi-1d case has been studied and evidence for a spin
polarization has been found \cite{Berggren98,Berggren96}. 
In Ref.~\onlinecite{Berggren98} the 
spin polarization in QPC's was studied at zero temperature assuming
that the expression in the local-density approximation for the
exchange energy of the bulk 2DEG is also valid in narrow quasi-1d
constrictions. This work suggested the possibility of spin
polarization in QPC's, but yielding only plateaus at 0.5 and 1.0, it
failed to explain the 0.7 anomaly. We base our phenomenological model
on these previous works on spin polarization. Without refering to any
particular model for the exchange energy, we simply assume the
existence on spin polarization expressed by a general spin-density
functional involving three unknown constants that are to be determined
experimentally. 

\begin{figure}
\centerline{\epsfysize=35mm\epsfbox{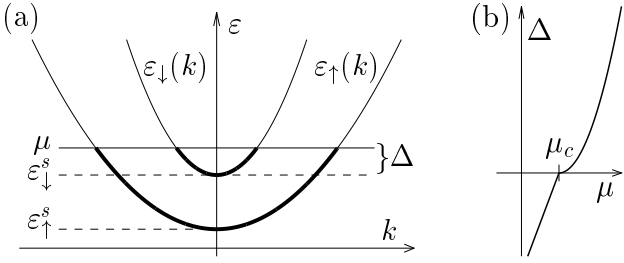}}
\narrowtext
\caption{ \label{fig:model}
(a) The instantaneous spin split subband structure of our model. (b)
The functional form Eqs.~(\ref{eq:Deltamu}) and
(\ref{eq:Deltamu_less}) of $\Delta(\mu) =
\mu-\varepsilon^s_\downarrow(\mu)$ giving rise to the 
anomalous 0.7 plateau.
}
\end{figure}

We stress that the spin polarization not necessarily is permanent. It
may well exhibit mesoscopic fluctuations. It is in fact enough to
assume that the dynamics of the spin degrees of freedom in the
constriction is adiabatic (slow) as compared to both the length of the
QPC and the time of passage of electrons through the QPC. Then, given
some instantaneous spin configuration the transmission coefficient
${\cal T}^{\rm tot}_{\sigma}$ for a spin-$\sigma$ electron going
through the QPC can thus be calculated as
${\cal T}^{\rm tot}_{\sigma} = {\cal T}_{\sigma}(E)
P_{\sigma} + {\cal T}_{\bar{\sigma}}(E) P_{\bar{\sigma}}$. Here
$P_{\sigma}$ ($P_{\bar{\sigma}}$) is the probability of finding
the incoming spin parallel (antiparallel) to the instantaneous
polarization. In the isotropic case with $P_{\sigma} =
P_{\bar{\sigma}}$ this leads to the same results as a static situation
where two spin subbands are formed. For simplicity we just treat the
case of a stationary polarization in the following and derive the
subband structure shown in Fig.~1a.

This discussion of spin-polarization serves as the justification of
the basic assumption in our work: the existence of a ``critical''
chemical potential $\mu_c$, where the cross-over from full to partial
polarization occurs. For $\mu < \mu_c$ the densities are $n_{\uparrow} =
n^0_\uparrow$ and $n_\downarrow = 0$, while for $\mu > \mu_c$ a
non-zero $n_\downarrow$ develops, i.e.\ two spin subbands appears with
different subband edges $\varepsilon_{\sigma}^s$ as depicted in Fig.1a.
It turns out that all the phenomenology of the anomalous plateaus is
contained in the $\mu$ dependence of the position of the minority
subband edge $\varepsilon_{\downarrow}^s(\mu)$ relative to $\mu$. This
important parameter is denoted $\Delta(\mu)$:
\begin{equation} \label{eq:Delta}
\Delta(\mu) = \mu-\varepsilon^s_\downarrow(\mu).
\end{equation}
Our entire analysis is based on the local spin density functional
written as
\begin{equation}
F=E[n_{\downarrow},n_{\uparrow}]-\mu\:(n_{\downarrow} +
n_{\uparrow}).
\end{equation}
Near the critical point $\mu_c$, for $\mu > \mu_c$,  we have
$n_{\downarrow} \ll n_{\uparrow}$ and the condition for the minimum of
the free energy becomes  
\begin{equation} \label{eq:free_energy}
\begin{array}{lclcl}
\frac{\partial F}{\partial n_{\uparrow}} & = &
\alpha + \alpha' \: 
\delta n_{\uparrow} + \gamma n_{\downarrow}-\mu&=&0, \\[2mm]
\frac{\partial F}{\partial n_{\downarrow}} & = &
\beta + \beta' n_{\downarrow}  + \gamma \: \delta n_{\uparrow}-\mu&=&0.
\end{array}
\end{equation}
We have made the linearization $n_{\uparrow} =
n_{\uparrow}^0 + \delta n_{\uparrow}$ for the majority spins and
assumed that the energy functional $F$ near the minimum is bilinear 
in $\delta n_{\uparrow}$ and $n_{\downarrow}$. 

The solution for the minority spin density in the case of $\mu >
\mu_c$ is $n_{\downarrow} \propto (\mu - \mu_c)$ which combined with
the 1d property that $n_{\downarrow}^2 \propto
\varepsilon_F^{\downarrow}$. But $\varepsilon_F^{\downarrow} =
\mu-\varepsilon_{\downarrow}^s = \Delta$, and we arrive at: 
\begin{equation}  \label{eq:Deltamu}
\Delta (\mu ) =
C(\mu -\mu _{c})^{2},\quad {\rm for\:} \mu > \mu_c.
\end{equation}
For $\mu <\mu _{c}$ we have $n_{\downarrow} = 0$ and $\Delta(\mu)$ is
now the excitation gap for flipping a spin at the Fermi level, i.e.\ 
$\Delta(\mu) = (\partial_{n_\uparrow}F-\partial_{n_\downarrow}F)=0$,
which gives $\Delta(\mu) = \beta + \gamma\: \delta n_{\uparrow}$. This
combined with $\delta n_\uparrow \propto \mu - \mu_c$ leads to:
\begin{equation}  \label{eq:Deltamu_less}
\Delta (\mu )= D(\mu -\mu _{c}),\quad {\rm for\:} \mu < \mu_c.
\end{equation}
We have thus derived the dispersion laws depicted in Fig.~1:
\begin{equation} \label{eq:dispersion}
\begin{array}{rcl}
\varepsilon_{\uparrow}(k) &=&
\frac{\hbar^2}{2m}k^2 + \varepsilon^s_{\uparrow},\\[2mm]
\varepsilon_{\downarrow}(k) &=&
\frac{\hbar^2}{2m}k^2 + \mu - \Delta(\mu).
\end{array}
\end{equation}

Given these dispersion relations, at finite temperature $T$ using an
idealized step-function transmission coefficient the LB conductance
$G(T)$ of this system is\cite{vanHouten92}
\begin{equation} \label{eq:dIdmu}
G(T)  = \frac{1}{2} G_2
\sum_{\sigma=\uparrow,\downarrow}
\int_{-\infty}^{\infty} d\varepsilon \:
\Theta(\varepsilon-\varepsilon^s_\sigma) \: 
\left(-f'[\varepsilon-\mu]\right),
\end{equation}
where $f'$ is the derivative of the Fermi-Dirac distribution
$f[x]=[\exp(x/k_BT)+1]^{-1}$ and $\Theta(x)$ is the step function.
By integration and using Eq.~(\ref{eq:Delta}) we obtain
\begin{equation} \label{eq:conductance}
G(T)= \frac{1}{2} G_2 \left(
f[\varepsilon^s_\uparrow-\mu] +
f[-\Delta(\mu)] \right).
\end{equation}
All predictions relating to experiments on the conductance of QPC's
follow from this simple analytical form. 

Before turning to a thorough
analysis of this expression we note some of its basic
features. Consider first the situation where the spin polarization is
nearly complete, i.e.\ $\Delta(\mu) \ll
\varepsilon^s_\downarrow(\mu)-\varepsilon^s_\uparrow(\mu)$. In this  
case, at low temperatures, $k_B T \ll \Delta(\mu)$,
both terms in Eq.~(\ref{eq:conductance}) are 1 and the conductance is
the usual $G_2$. However, in the temperature range
\begin{equation} \label{eq:condition}
\Delta(\mu) \ll k_B T \ll
\varepsilon^s_\downarrow(\mu)-\varepsilon^s_\uparrow(\mu),
\end{equation}
the contribution of the first term is 0.5 while the second term
remains 1 yielding $G = 0.75\: G_2$. Due to the parabolic
$\mu$-dependence of $\Delta$ given by Eq.~(\ref{eq:Deltamu}), which in a
sense pins $\varepsilon_{\downarrow}^{s}$ to $\mu$, the condition
(\ref{eq:condition}) is in fact fulfilled for a sufficiently broad 
range of $\mu$ (in experiments $\mu \propto V_g$), thus giving rise to
a 0.7 quasi-plateau.

An in-plane magnetic field {\bf B} is readily taken into account
by adding Zeeman energy terms and substituting
\begin{equation} \label{eq:zeeman}
\varepsilon_{\uparrow}^s \rightarrow
\varepsilon^s_{\uparrow}-g\: \mu_B |{\bf B}|, \qquad
\varepsilon^s_{\downarrow} \rightarrow
\varepsilon^s_{\downarrow}+g\: \mu_B |{\bf B}|.
\end{equation}

\section{Experimental implications of the model}
In the following we
discuss how the model can explain the experimental observations
(e1)-(e6). In Fig.~2(a) observations (e1) and (e2) are clearly seen in
the model calculation. In this idealized case with a
step-function transmission coefficient the plateau appears
at $0.75$ as discussed above.
\begin{figure}
\centerline{\epsfysize=40mm\epsfbox{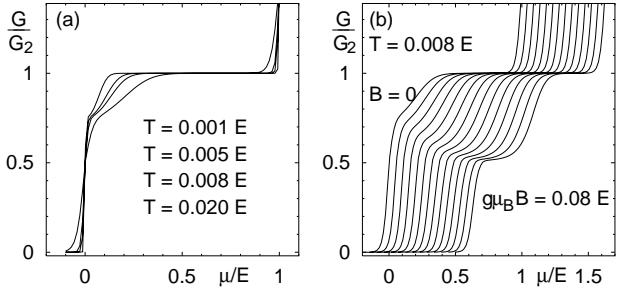}}
\narrowtext
\caption{\label{fig:GT}
(a) The conductance from
Eqs.~(\protect\ref{eq:conductance}) and~(\protect\ref{eq:Deltamu})
with $C=0.5$, $D=1.0$ and $\mu_c=\varepsilon^s_\uparrow$. All energies are
given in units of the transverse mode subband spacing $E$.
(b) The conductance as in (a) but with fixed $T=0.008~E$ and an applied
in-plane magnetic field varying from 0 to 0.08~$E/g\mu_B$. For clarity the
curves in (b) have been offset horizontally.
}
\end{figure}
Observation (e3) follows trivially from Eq.~(\ref{eq:dIdmu}) with
the activation temperature $T_a = \Delta(\mu)$. Assuming that in
the vicinity of $\mu_c$ the chemical potential depends linearly on
the gate voltage $V_g$ Eq.~(\ref{eq:Deltamu}) immediately predicts
(e4) with the exponent $\alpha=2$.

The result of the model calculation in the presence of a magnetic
field using Eqs.~(\ref{eq:conductance}), (\ref{eq:Deltamu})
and~(\ref{eq:zeeman}) is shown in Fig.~2(b). 
In accordance with observation (e5) the 0.7
anomaly develops smoothly into an ordinary Zeeman split 0.5
plateau. The experimental observation (e6) concerns finite
bias. This brings us into a strong non-equilibrium situation which
is outside the scope of the present work. However, considering a
small finite bias not too far from the equilibrium case, we
do find that the 0.75 plateau rises, which gives
additional support for the picture presented here. In
Ref.~\cite{Kristensen00} the finite bias measurements were well
explained by straightforward extension of our model to the non-linear
regime.

\begin{figure}
\centerline{\epsfysize=40mm\epsfbox{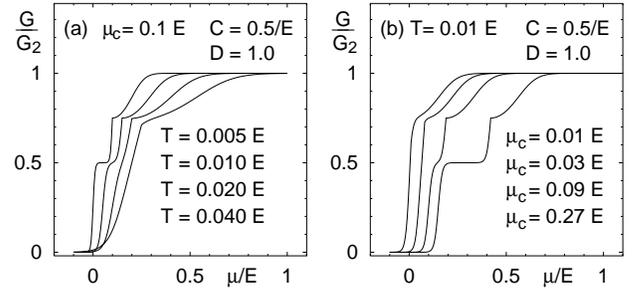}}
\narrowtext
\caption{\label{fig:Tmuc}
(a) $G$ for fixed $\mu_c$ and coefficient $C$ while changing
temperature. For clarity, the curves are offset, the temperature
increasing from left to right. Note the disappearence of the 0.5
plateau at high $T$. (b) $G$ for fixed temperature and coefficient $C$
while changing $\mu_c$. For clarity, the curves are offset, $\mu_c$ 
increasing from left to right. Note the disappearence of the 0.5
plateau for small values of $\mu_c$.
}
\end{figure}

Apart from 0.7 quasiplateau, our model also predicts a
plateau at $0.5\: G_2$, which should be seen in
perfect QW/PCs at zero temperature and zero
magnetic field when $\varepsilon^s_\uparrow < \mu < \mu_c$ (full
polarization). In order to observe the 0.5 plateau the condition 
$k_BT < \min\{|\Delta|, \mu - \varepsilon^s_\uparrow\}$ 
must be fulfilled. Otherwise thermal smearing will destroy the
plateau. Note that even if the 0.5 plateau is thermally smeared,
Eq.~(\ref{eq:condition}) may still hold and result in a 0.7 plateau. 
This is illustrated in Fig.~3. In Fig.~3(a) $\mu_c$ is fixed and the
temperature is changed, while in Fig.~3(b) the value of $\mu_c$ is
changed at a fixed temperature. 
One can see that in a situation where $T<\mu_c/8$ the 0.5 and the 0.7
plateaus can be seen simultaneously. This situation has never been
observed experimentally, which might be an indication that the
experimentally accessible temperatures are too high to resolve the 0.5
plateau. However according to (e7), a situation where $0.5$ plateau
appeared without a $0.7$ plateau was observed. This might happen if
the range of $\mu$ corresponding to partial polarization 
in the QPC  was particularly narrow as compared to that
corresponding to full polarization. In terms of
the parameters of our model it would mean e.g.\ a very
large constant $C$. 

The role of parameter $C$ is illustrated
in Fig.~4. One can see that with increasing $C$ the 0.7
quasiplateau becomes narrower and less prominent, and practically
disappears in comparison with the $0.5$ at very large values of
$C.$

\section{Non-ideal transmission}

In our idealized model we used a step-function transmission
coefficient for electrons traversing the QPC. Non-ideal
transmission is easily taken into account by replacing the theta
function in Eq.~(\ref{eq:dIdmu}) with a given transmission coefficient 
${\cal T}_{\sigma}(\varepsilon)$:
\begin{equation} \label{eq:dIdmuT}
G(T)  = \frac{1}{2} G_2
\sum_{\sigma=\uparrow,\downarrow}
\int_{-\infty}^{\infty} d\varepsilon \:
{\cal T}_{\sigma}(\varepsilon) \: 
\left( -f'[\varepsilon-\mu] \right).
\end{equation}
Through the transmission coefficient
${\cal T}_{\sigma}(\varepsilon)$
the conductance now depends on the geometry of the QPC
and is no more universal. However, some qualitative predictions
can still be made without a knowledge of the exact transmission
properties of the QPC.
First of all, due to the conditions Eqs.~(\ref{eq:condition})
and~(\ref{eq:Deltamu}) the quasi-plateau
persists. Secondly, mainly the transmission coefficient
of minority spin band is affected which results in a suppression
of the anomalous plateau while the integer plateau remains close
to 1. Model calculations show that the anomalous 0.7 plateau
may be suppressed to the values of $ G=0.6 G_2 $ without
being destroyed \cite{BCF}.
Another effect of non-ideal transmission is the "smearing" of the
$G$ vs. $\mu$ dependence as compared to the situation of ideal
transmission. The character of smearing is not unlike the thermal
smearing, and it may e.g.\ contribute to the effect of
masking the $0.5$ plateau in the samples where only the
$0.7$ plateau is seen.

\begin{figure}
\centerline{\epsfysize=40mm\epsfbox{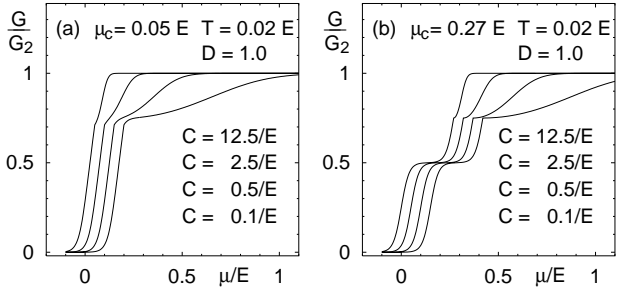}}
\narrowtext
\caption{\label{fig:C}
(a) $G$ for fixed temperature and a small value of $\mu_c$ while
changing the coefficient $C$. For clarity, the curves are offset, $C$
decreasing from left to right. Note the emerging 0.7 plateau at small
$C$. (b) The same as in (a) except for a larger value of $\mu_c$. Note
the coexistence of the 0.5 and 0.7 plateaus.
}
\end{figure}

\section{Summary and conclusion}
We have presented a phenomenological model which can
account for the experimental observations of the anomalous 0.7 and
0.5 conductance plateaus in mesoscopic QPCs. The model is built on an
assumption of an effective instantaneous partial polarization seen by
the traversing electrons, while the ground state itself needs not
have a finite magnetic moment. Even in its simplest form excluding
effects of non-ideal transmission our model is capable of
providing a very satisfactory description of different experimental
situations, which is achieved by varying the two model parameters
$\mu_c$ and $C.$ 

\subsection*{Acknowledgements}
The authors acknowledge support from the Danish Natural
Science Research Council through Ole R{\o}mer Grant No.\ 9600548.


\end{document}